# Beyond Answers: Large Language Model-Powered Tutoring System in Physics Education for Deep Learning and Precise Understanding

Zhoumingju Jiang*

School of Design, Southern University of Science and Technology, Shenzhen, China, jiang2020@sustech.edu.cn

Mengjun Jiang*

Hongshan Middle School, Shenzhen, China, 763534670@qq.com

The integration of artificial intelligence (AI) in education has shown significant promise, yet the effective personalization of learning, particularly in physics education, remains a challenge. This paper proposes Physics-STAR, a framework for large language model (LLM)-powered tutoring system designed to address this gap by providing personalized and adaptive learning experiences for high school students. Our study evaluates Physics-STAR against traditional teacher-led lectures and generic LLM tutoring through a controlled experiment with 12 high school sophomores. Results showed that Physics-STAR increased students' average scores and efficiency on conceptual, computational, and on informational questions. In particular, students' average scores on complex information problems increased by 100% and their efficiency increased by 5.95%. By facilitating step-by-step guidance and reflective learning, Physics-STAR helps students develop critical thinking skills and a robust comprehension of abstract concepts. The findings underscore the potential of AI-driven personalized tutoring systems to transform physics education. As LLM continues to advance, the future of student-centered AI in education looks promising, with the potential to significantly improve learning outcomes and efficiency.

CCS CONCEPTS • **Human-centered computing → Interaction design theory, concepts and paradigms; • Computing methodologies → Artificial intelligence; • Applied computing → Education;**

## 1 INTRODUCTION

Personalized support for students is a gold standard in education, yet it implemented poorly with the large number of students[8]. To address this challenge, intelligent tutoring systems(ITS) have been developed to engage the students in sustained reasoning activity, and interact with student based on their needs[6, 19]. However, these systems are expensive and time-consuming to build. AI-based intelligent tutoring systems, such as large language models(LLMs), have

* Both authors contributed equally to this research

demonstrated potential for personalized tutoring in subjects like writing, biology and computer science[1, 9, 12]. Yet, research in physics education remains limited. Physics education has unique requirements, such as the need for abstract thinking, model construction skills, and the ability to apply mathematical models to solve problems[18]. Current LLM-assisted systems often fall short of these specific requirements, leading to seemingly correct but wrong guidance: misinterpretation of concepts, erroneous calculation, and incorrectly extracting information from diagrams. While the rapid iteration of LLM models (e.g., GPT-4o and Gemini) offers an opportunity to address these issues, advancements in AI capabilities have not yet translated into significant improvements in student capabilities.

To bridge the gap between advanced AI capabilities and student needs in physics education, we propose a LLM-powered tutoring system designed to facilitate deep learning for students. This system aims to help students learn from their mistakes, understand physics concepts and models accurately, and internalize this knowledge. Additionally, it enhances the learning experience through flexible, personalized questions and answers. We also present preliminary results from an evaluation conducted with a group of 12 high school sophomores. The primary goal of this study is to introduce an adaptable framework for guiding the LLM to provide personalized tutoring in physics education. This framework aims to be scalable for future experiments testing hypotheses about the role of LLM-powered tutoring system in personalized learning environments.

## 2 RELATED WORKS

Intelligent tutoring systems(ITS) have been employed to enhance personalized learning in education. Previous work in the intersection of human-computer interaction (HCI), computer-supported cooperative work (CSCW), and education has explored how ITS can improve educational utility and learning performance through tutorial dialogue[5, 10, 11]. Notable examples include AutoTutor[10], a conversational tutoring software that aids learning in physics, computer science by fostering deep understanding and reflective thinking. Similarly, Heffernan et al. developed the ASSISTments system, integrating intelligent tutoring with assessment features, offering real-time feedback to students and data-driven reports to teachers during homework activities[11]. Although these systems highlighted the potential of ITS in personalized learning, they were expensive and time-consuming to build.

Large language models enhance the interaction between ITS and students by facilitating ideas exchanges using natural language in a cost-effective manner. For example, Kumar et al. explored how AI-driven instructional strategies can improve interactions between students and LLM assistants to achieve satisfied learning outcomes[14]. Cai et al. developed chatbots based on collaborative learning theories, allowing students to benefit from guided interactions with LLM[4]. Additionally, Huang et al. investigated the potential of LLM to improve learning outcomes by providing personalized scaffolding and feedback[13]. These studies underscored the promising role of LLM in enhancing educational engagement and effectiveness.

Despite these advancements, the use of LLMs in personalized learning environments, particularly in physics education, still faces significant challenges. The application of LLM is limited by their inadequate physical knowledge and restricted ability to extract complex information. For example, Yeadon and Halliday found that ChatGPT could generate seemingly plausible but incorrect answers in physics exams[21], potentially misleading students. Similarly, Dahlkemper et al. highlighted the necessity for caution when relying on AI-generated responses[7]. Recent research has aimed to improve the accuracy and computational power of LLM information retrieval through methods such as prompt engineering[15–17], fine-tuning[2, 3], and knowledge graph[20, 22]. However, these efforts primarily focus on knowledge acquisition and problem-solving skills of LLMs, neglecting the needs of students in personalized learning, such as analyzing specific errors and applying knowledge for deeper understanding. To address these limitations, we propose the Physics-STAR framework, which utilizes prompt engineering to generate LLM inputs tailored for personalized physics learning environments.

## 3 METHOD

### 3.1 Framework of Physic-STAR

The framework of Physics-STAR, drawing on the Situation-Task-Action-Result(STAR) approach, provides an adaptable structure for LLM input in three steps of personalized tutoring, as shown in Fig. 1. The usual process of the revision course is divided into three stages. The first stage, "quiz", is a test to understand the weak points in the student's knowledge; In the second stage, "tutoring", the teacher provides a lecture on the topics with a high error rate; Finally, in "practice" stage, the teacher conducts a lecture to evaluate whether the students have mastered the knowledge points, thus assessing the teaching effectiveness.

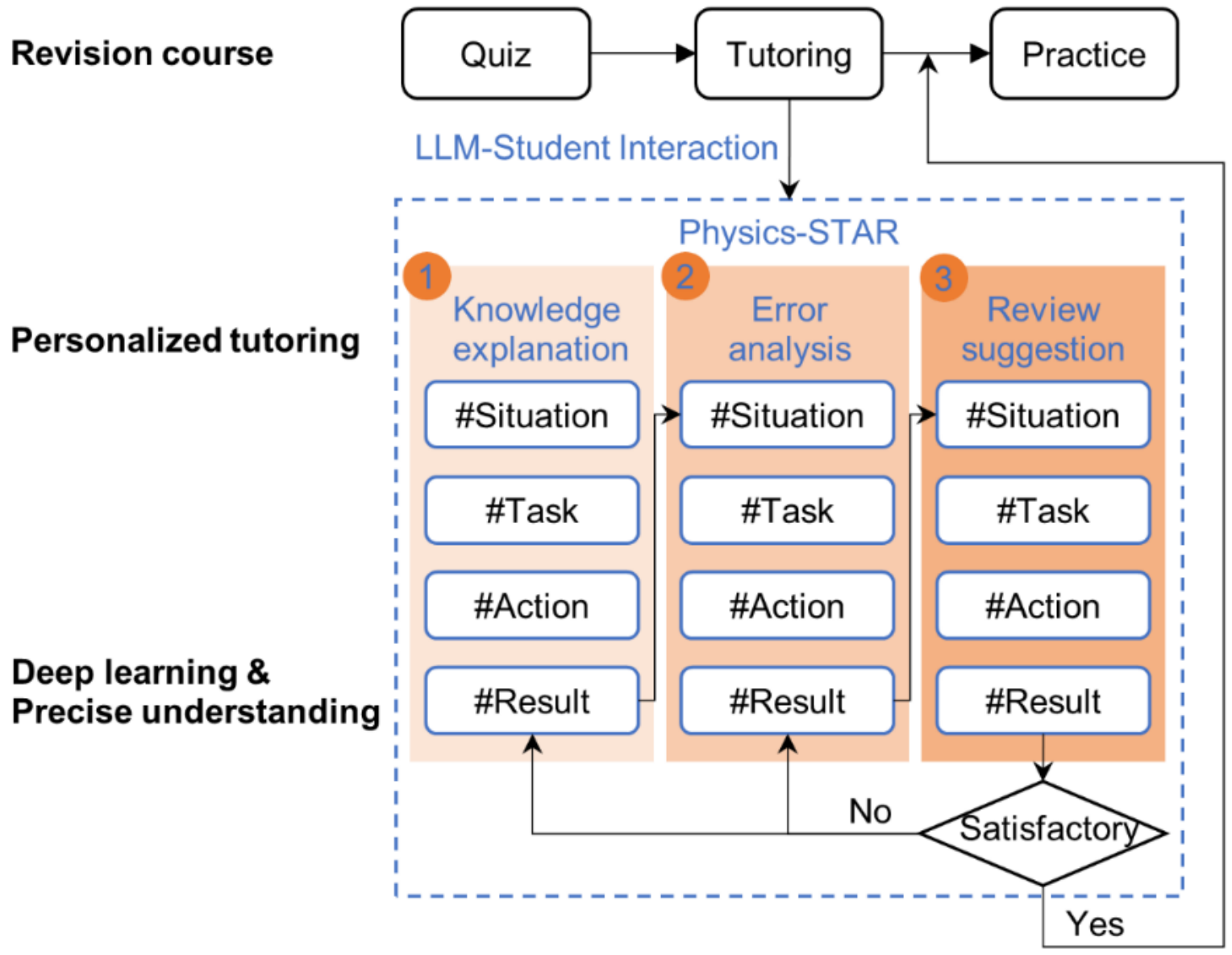

Figure 1: The framework of Physics-STAR. Physics-STAR supports students in deep learning and precise understanding of physics knowledge through personalized tutoring. During LLM-student interaction, the LLM provides tailored guidance based on the student's asking prompts in three steps: knowledge explanation, error analysis, and review suggestion. In each step: the LLM prompt is designed as the Physics-STAR, including situation, task, action, and results.

In the traditional context of the teacher lectures, the revision course is characterized by unilateral output from the teacher to students, with no information exchange. This approach can only address the common problems for the entire class and fails to provide personalized guidance to individual students, resulting in suboptimal learning outcomes and limited development of students' deep learning and precise understanding. A personalized tutoring system tailored for individual students can better address knowledge gaps and enhance deep learning. Based on LLM, we propose the Physics-STAR framework, which consists of three steps:

1. Knowledge Explanation: This step involves explaining the concepts, basic formulas, and application scenarios of specific knowledge to students.

2. Error Analysis: Through the communication with the student, this step analyzes the reasons behind student's mistakes on a certain questions.

3. Review Suggestion: Personalized review suggestions are provided based on the analysis of errors. Following this, similar questions are presented to the students for testing. If they pass the test, it indicates that the knowledge point has been adequately understood; otherwise, the three-step process is repeated. The structure of prompts in each step consists of four sections: a situation, a task, an action, and a result, as shown in Table 1. Students are required to engage in question-and-answer sessions with the personalized tutoring system according to the prompts provided.xxq

Table 1 The prompt of Physics-STAR used to guide LLM provides knowledge explanation, error analysis, and review suggestions.

| **Prompt 1:** | **For knowledge explanation** |
|---|---|
| Student Prompt: | #Situation: describe the question's context and options<br>#Task: explain the question's basic knowledge points, formulas, and solution methods<br>#Action: provide a reference solution process<br>#Result: hints for correct answers |
| GPT-4o: | [Knowledge explanation]<br>(1)Basic knowledge explanation: [principles and formulas]<br>(2)Solution process: [option analysis]<br>(3)Correct Answer: [specific option] |
| **Prompt 2:** | **For error analysis** |
| Student Prompt: | #Situation: description of student's problem-solving process with errors<br>#Task: analyze the reason for the error<br>#Action: remind the correct understanding and formula to be used at the error<br>#Result: correct the error |
| GPT-4o: | [Error analysis]<br>(1)Analyzing the causes of student errors<br>(2)Remind the errors and correct understanding<br>(3)Correct understanding and calculation formula<br>(4)Conclusion: emphasize the cause of the error |
| **Prompt 3:** | **For review suggestion** |
| Student Prompt: | #Situation: students summarize and reflect on why topics are wrong<br>#Task: provide advice on revision guidelines for students<br>#Action: review the basic formulas, redo the questions, and practice more questions<br>#Result: assess the correct rate of similar questions |
| GPT-4o: | [Review suggestion]<br>(1)Review the basic formulas<br>(2)Remind the errors and correct understanding<br>(3)Provide three similar questions:<br>[question1], [question 2], [question3]<br>(4)Assess the correct rate of the similar questions,<br>If the correct rate is high: [The student has mastered the knowledge]<br>If the correct rate is low: [repeat the prompts 2 and 3 ] |

### 3.2 Participants and Experiment Setting

Twelve high school sophomores(six males and six females) were divided into three groups of four students each. Each student's usual physics score is similar (between 60-70 out of 100). The first group was taught using a traditional teacher lecture method without LLM intervention, the second group used a general LLM(GPT-4o), and the third group utilized GPT-4o with the guidance of Physics-STAR. None of the 12 students had any prior experience with LLMs.

This experiment employed a control group pretest-posttest design to compare the performance of students using the LLM (experimental group) and those not using the LLM (control group) before and after the experiment. Additionally, this design compared the efficacy of different tutoring strategies within the LLM: the generic LLM prompting strategy(control group) versus the Physics-STAR prompting strategy (experimental group). The experimental process is illustrated in Fig. 2.

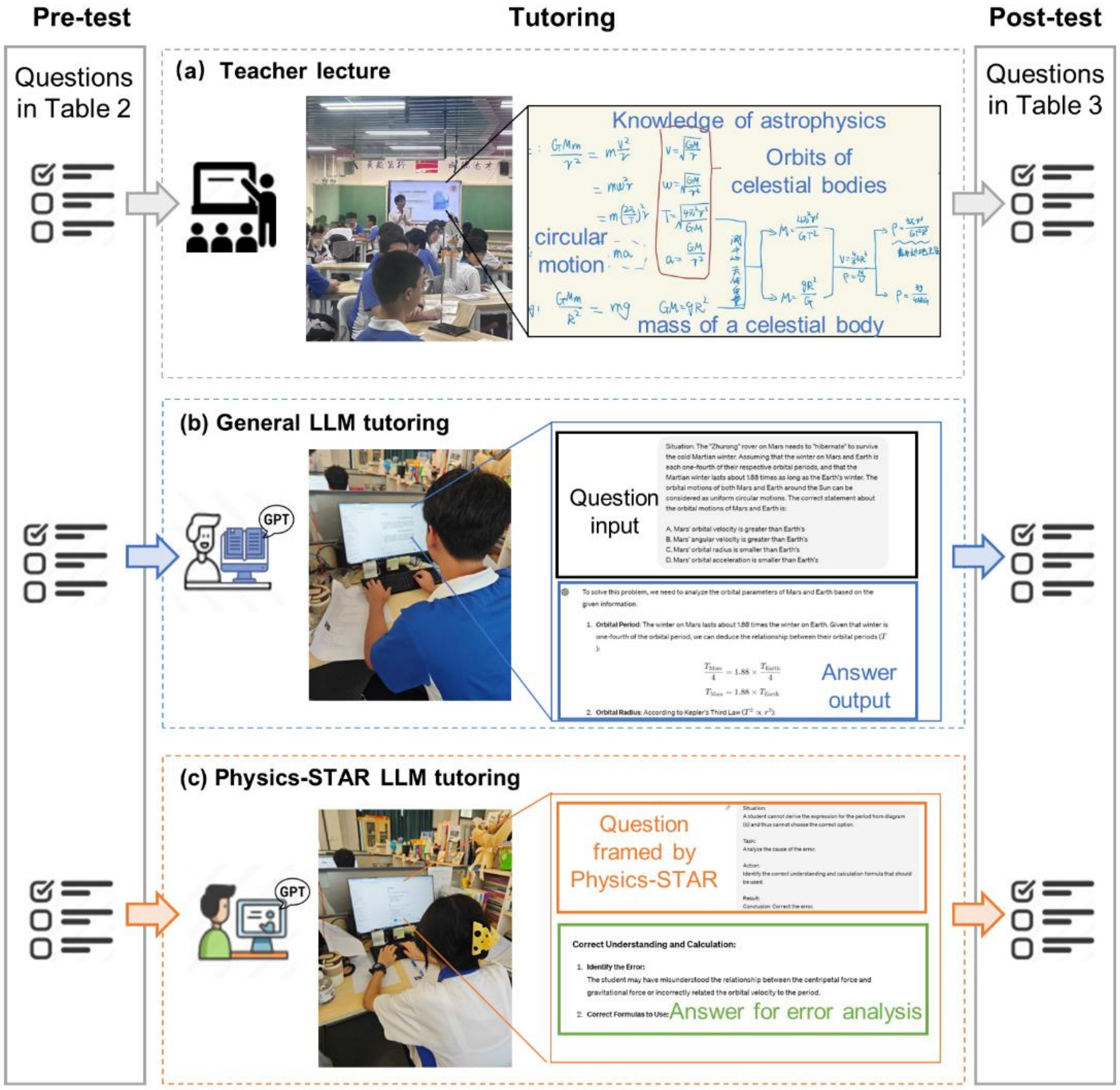

Figure 2: The experiment process comparing three tutoring approaches: (a) teacher tutoring in class, (b) general LLM tutoring, and (c) Physics-STAR LLM tutoring. We conducted the experiment based on the review class of astrophysics in high school physics knowledge, selected 12 eleventh-grade students as the experimental subjects, and divided into three groups using different methods to review the content.

A pre-test was administered to all participating students to gather baseline data prior to the start of the experiment. Each of the three groups then engaged with their respective tutoring strategies: the first group received traditional physics review lectures, the second group interacted freely with GPT-4o for review, and the third group interacted with GPT-4o guided by the Physics-STAR framework. A post-test was administered at the end of the experiment to evaluate student learning outcomes. Students were assessed on their understanding of specific high school physics concepts related to gravity and spaceflight.

The pre-test and post-test each consisted of three types of multiple-choice questions: conceptual, computational, and informational, worth 10 points each, for a total of 30 points. The purpose of these tests was to determine whether participants had a comprehensive understanding of the law of gravity and spaceflight. The details of the tests are shown in Table 2 and Table 3 in the Appendices. Conceptual questions assessed basic knowledge, such as recalling the formula for the law of gravity. Computational questions evaluated both knowledge and computational skills, such as applying the law of gravity to analyze the motion of celestial bodies in specific scenarios. Informational questions were designed to assess abstract thinking and numeracy, such as interpreting physical information from a diagram and analyzing the motion of a celestial body in relation to the law of gravity.

## 4 RESULT

Our experiment evaluated student performance using pre-tests and post-tests across three groups: teacher lecture, general LLM tutoring, and Physics-STAR LLM tutoring. The scores and time taken for each question of pre-test and post-test are shown in Fig. 3.

### 4.1 Pre-Test Analysis

In the pre-test, the overall scores and completion times were comparable across the three groups. Conceptual questions had the highest average score (M = 8.33) and the shortest completion time (M = 54.42s). Computational questions had moderate scores (M = 5.83) and completion times (M = 119.17s). Information-based questions were the most challenging, with the lowest average scores (M = 1.67) and the longest completion times (M = 273.83s).

### 4.2 Post-Test Analysis

*Between-Group Analysis.* For Conceptual Questions, all groups achieved perfect scores. Completion times were similar: Teacher lecture (M = 49.5s), General LLM tutoring(M = 50.25s), and Physics-STAR LLM tutoring (M = 53s). For Computational Questions, scores were consistent across groups, with 75% of students achieving full marks. Completion times were close: Teacher lecture (M = 113.5s), General LLM tutoring (M = 111s), and Physics-STAR LLM tutoring (M = 108.25s). For Information-Based Questions, the Physics-STAR group scored highest (M = 5), with completion times (M = 257s). General LLM tutoring and Teacher lecture groups had similar scores (M = 2.5) with times of (M = 262s) and (M = 271.5s), respectively.

*Within-Group Analysis.* For teacher teaching, scores improved similarly for conceptual (M = 10) and computational questions (M = 7.5), with limited improvement in information-based questions (M = 2.5). For General LLM tutoring, scores for conceptual and computational questions were identical to the Teacher lecture, with information-based questions also showing minimal improvement. For Physics-STAR LLM tutoring, similar improvements were observed for conceptual and computational questions, with a more significant improvement in information-based questions, increasing the number of correct responses from one to two.

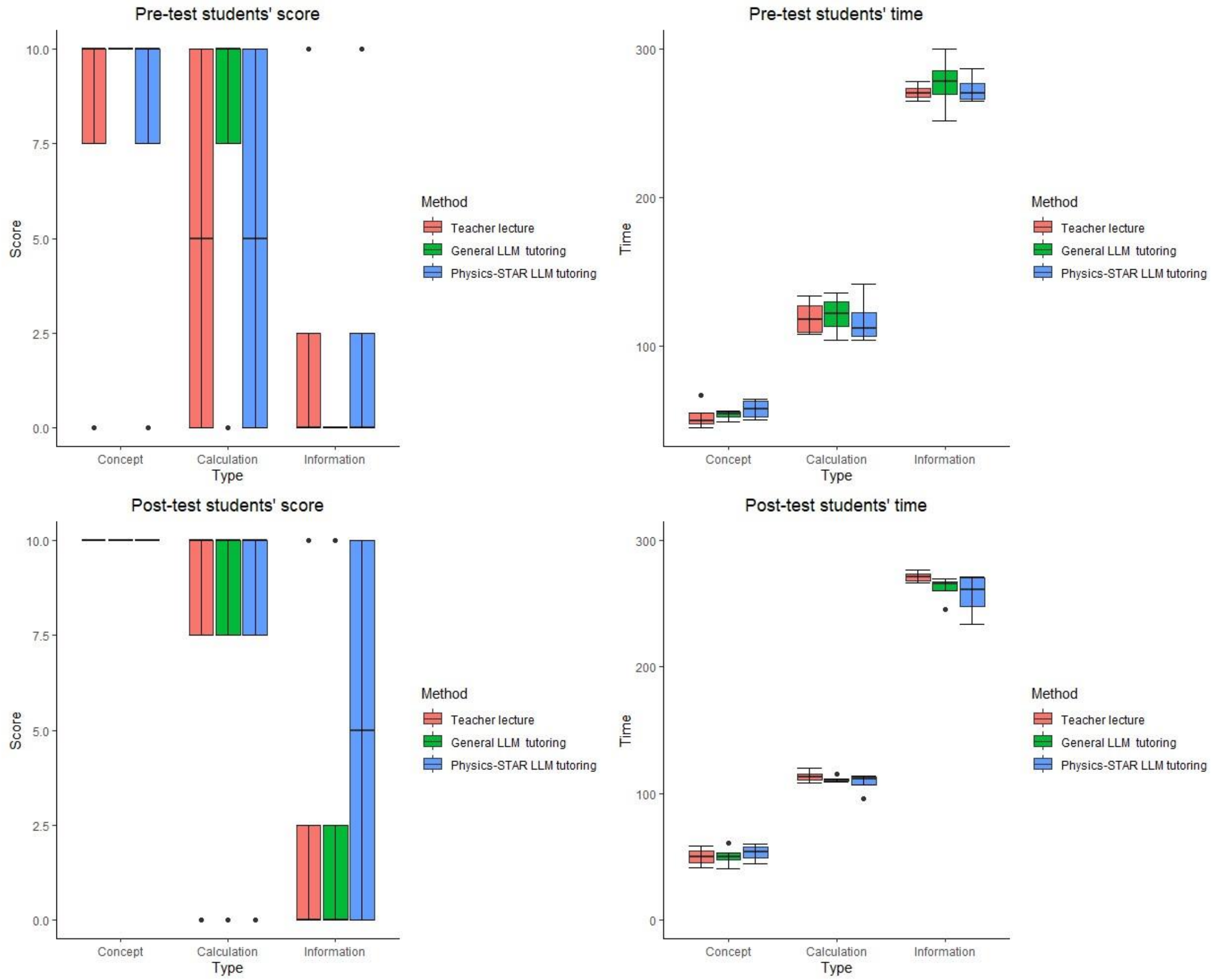

Figure 3: Results of the test score. Small dots refer to individual student's grades, while lines represent the average value.

### 4.3 Analysis of Score and Time Differences

*Between-Group Analysis.* Improvements in conceptual and computational question scores were consistent across teaching methods. However, the Physics-STAR method showed significantly better improvement in information-based questions. Time spent on conceptual and computational questions remained similar between pre-tests and post-tests, with reduced variance in completion times for computational questions. The Physics-STAR method showed a notable decrease in time spent on information-based questions compared to the other two groups.

*Within-Group Analysis.* Teacher lecture: Minimal differences in completion times for all question types between pre- and post-tests, with decreased variance in computational questions. General LLM tutoring: Stable completion times for conceptual questions; reduced times for computational and information-based questions by approximately 10 seconds, with decreased variance. Physics-STAR LLM tutoring: Similar reductions in time for conceptual and computational questions, with a more pronounced decrease for information-based questions.

Overall, students using the Physics-STAR LLM tutoring scored higher on understanding assessments and demonstrated greater efficiency in reasoning tasks, particularly in addressing complex information-based questions. Our results reveal distinct impacts of the three instructional methods: teacher lecture, general LLM tutoring, and Physics-STAR LLM tutoring on various question types. For conceptual questions, all methods equally reinforced students' understanding of fundamental concepts, indicating that basic conceptual comprehension can be effectively achieved through any of these teaching approaches. In computational questions, the demonstration-based teaching methods used by all three groups allowed students to replicate problem-solving techniques without deeply engaging their cognitive processes. This uniformity suggests that in-depth interaction during teaching does not significantly influence computational task outcomes.

Notably, for information-based questions, where success depends on accurately interpreting and deeply reflecting on the problem's information, the Physics-STAR LLM tutoring method stood out. By engaging students in interactive Q&A sessions, Physics-STAR LLM tutoring helped students understand their errors and enhance their analytical thinking skills. This interactive approach led to more stable and improved performance on information-based questions compared to teacher lectures and general LLM tutoring, which did not facilitate the same level of deep cognitive engagement.

## 5 DISCUSSION

### 5.1 Prioritizing Student-Centered Approaches in Physics Education with Large Language Model

Physics education must be fundamentally student-centered rather than AI-centered to ensure meaningful educational outcomes. While LLM offers significant advantages in terms of accessibility, efficiency, and scalability, their primary function should be to support and enhance the learning process tailored to each student's unique needs and learning style. Current LLM-powered tutoring systems often prioritize the capabilities of AI over the nuanced requirements of students. This misalignment can lead to a superficial understanding of physics concepts, as these systems tend to focus on problem-solving capabilities without fostering deep comprehension or critical thinking skills. To truly benefit students, LLM must be designed to facilitate active learning, encouraging students to engage deeply with the material, visualize complex ideas, and develop problem-solving strategies through guided exploration rather than mere answer retrieval.

Furthermore, personalized learning in physics requires LLM-powered tutoring systems that can adapt to the varying levels of students' prior knowledge and cognitive abilities. This involves providing contextualized feedback and scaffolding that helps students build a strong foundational understanding before progressing to more complex topics. The role of LLM should be to complement human teachers by offering individualized support that teachers, constrained by time and resources, might not be able to provide. By shifting the focus from LLM's problem-solving prowess to its potential as a personalized learning aid, we can create more effective educational environments that prioritize student engagement, understanding, and long-term retention of physics principles.

### 5.2 Enhancing Error Analysis and Reflective Learning in LLM-Powered Physics Education

A critical limitation of current LLM-powered tutoring systems in physics education is their tendency to provide direct answer retrieval without offering an in-depth analysis of the causes of errors. This approach deprives students of valuable learning opportunities that arise from understanding and correcting their mistakes. Effective learning, particularly in a subject as conceptually demanding as physics, requires students to engage in reflective practice, where they not only receive the correct answers but also comprehend the underlying principles and reasoning that lead to those answers. By focusing solely on providing immediate solutions, LLM-powered systems risk promoting a superficial grasp of the material, where students may memorize correct responses without internalizing the processes needed to arrive at them.

To address this issue, LLM-powered tutoring systems must evolve to incorporate features that emphasize error analysis and critical thinking. For instance, instead of simply presenting the correct answer, AI could guide students through a step-by-step breakdown of their errors, highlighting misconceptions and providing explanations that reinforce conceptual understanding. This would transform mistakes into learning moments, encouraging students to analyze their problem-solving strategies and develop a deeper comprehension of physics concepts. Additionally, these systems should be designed to prompt students to reflect on why a particular answer is correct or incorrect, fostering an inquisitive mindset and enhancing their ability to apply knowledge in varied contexts.

### 5.3 Balancing AI Integration with Mastery of Fundamental Skills

The rapid advances in AI technology in education have raised concerns about whether students are neglecting the basics of foundational knowledge. While LLM offer significant advantages in terms of efficiency and personalized learning, they can inadvertently encourage students to rely too heavily on these technologies for problem-solving. This dependency risks diminishing students' engagement with fundamental concepts and principles, which are crucial for a deep understanding of subjects like physics. LLM often designed to provide quick and accurate answers, may lead students to focus more on obtaining the correct results rather than understanding the underlying processes and theories.

To counteract this potential issue, it is imperative that AI educational tools are designed to reinforce the basics rather than replace them. AI should be used to augmentTeacher lecture methods by providing supplementary practice that focuses on fundamental skills and concepts. For example, AI can identify areas where a student struggles with the basics and offer targeted exercises that reinforce these areas. Additionally, AI tools should encourage students to manually work through problems before turning to automated solutions, ensuring that they engage deeply with the material and develop a robust foundational understanding. By striking a balance between leveraging AI's capabilities and maintaining a strong emphasis on the basics, we can ensure that students benefit from technological advancements without compromising their mastery of essential skills.

## 6 CONCLUSION

In this research, we propose an LLM-powered tutoring framework for physics education: Physics-STAR, a practical framework designed to guide LLMs to generate personalized feedback for individual students. This framework aims to meet the challenges in personalized tutoring and bridge the gaps between LLM capacity and student needs in solving physics problems. The experiment demonstrates Physics-STAR's effectiveness in enhancing students' precise understanding and deep learning of physics concepts, particularly in its ability to closely align with students' needs.By guiding students step-by-step through abstract problems and extracting information from real-world scenarios, Physics-STAR not only improves student performance in physics but also provides an effective way to integrate LLMs into the teaching of various disciplines. As LLM technology continues to evolve, multimodal knowledge interactions will further advance student-centered AI in education, enhancing the learning experience and ensuring that AI tools support meaningful and personalized educational outcomes.

## ACKNOWLEDGMENTS

We would like to thank the 12 students at Shenzhen Hongshan Middle School. We thank the reviewers for their feedback on the paper, Paul and Dr.Adrian Rowland, who helped develop the framework.

## A APPENDICES

Table 2 Sample Pretest Questions

| | |
|---|---|
| Question Type 1: Conceptual Questions | |
| Question: | In April 2021, China's independently developed space station core module "Tianhe" was successfully launched and entered orbit. If the core module's orbit around the Earth is considered uniform circular motion, and given the gravitational constant, which of the following physical quantities can be used to calculate the Earth's mass? |
| Options: | A. The mass of the core module and the radius of the orbit<br>B. The mass of the core module and the orbital period<br>C. The angular velocity of the core module and the orbital period<br>D. The orbital speed of the core module and the radius of the orbit |
| Correct Answer: | D |
| Question Type 2: Computational Questions | |
| Question: | Mars has a diameter about half that of Earth and a mass about one-tenth that of Earth. Its orbital radius around the Sun is about 1.5 times that of Earth's orbit. Based on this data, which of the following statements is correct? |
| Options: | A. The gravitational acceleration on the surface of Mars is smaller than that on Earth<br>B. Mars' orbital period around the Sun is shorter than Earth's<br>C. Mars' orbital speed around the Sun is greater than Earth's<br>D. The centripetal acceleration of Mars in its orbit around the Sun is greater than Earth's |
| Correct Answer: | A |
| Question Type 3: Information-Based Questions | |
| Question: | As shown in figure (a), an exoplanet P orbits a star Q in uniform circular motion. Due to the occlusion by P, the brightness of Q detected by a probe varies periodically with time, as shown in figure (b), with the same period as P's orbit. Given that the mass of Q is known and the gravitational constant is G, which of the following statements about P's orbit is correct? |
| Figure: | 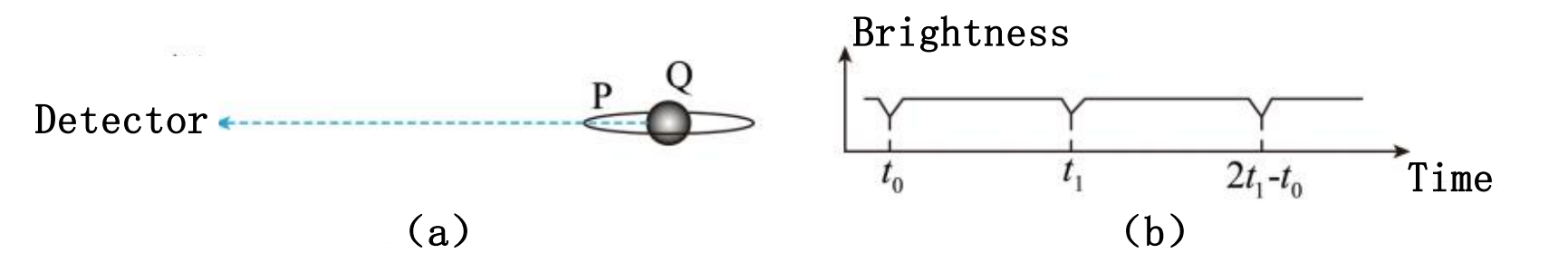 |
| Options: | A. $T = 2t_1 - t_0$<br>B. Orbital radius. $r=\sqrt[3]{\frac{GM\left(t_1-t_0\right)^2}{4\pi^2}}$<br>C. Angular velocity. $\omega = \frac{\pi}{t_1 - t_0}$<br>D. Acceleration. $a=\sqrt[3]{\frac{2\pi GM}{t_1-t_0}}$ |
| Correct Answer: | B |

Table 3 Sample post-test questions

| | |
|---|---|
| **Question Type 1: Conceptual Questions** | |
| Question: | Jupiter is one of the planets orbiting the Sun, and it has satellites orbiting around it. To determine the mass of Jupiter through observation, which quantities need to be measured? (Given the universal gravitational constant *G*) |
| Options: | A. Jupiter's orbital period around the Sun and its orbital radius<br>B. Jupiter's orbital period around the Sun and its orbital speed<br>C. The satellite's orbital period around Jupiter and Jupiter's radius<br>D. The satellite's orbital period around Jupiter and its orbital radius |
| Correct Answer: | D |
| **Question Type 2: Computational Questions** | |
| Question: | The "Zhurong" Mars rover needs to "hibernate" to survive the cold winter on Mars. Assuming that the winter on Mars and Earth each lasts one-quarter of their respective orbital periods, and Mars' winter duration is approximately 1.88 times that of Earth's winter. Both Mars and Earth's orbits around the Sun can be considered uniform circular motion. Which of the following statements about the orbits of Mars and Earth is correct? |
| Options: | A. Mars' orbital speed is greater than Earth's<br>B. Mars' angular velocity is greater than Earth's<br>C. Mars' orbital radius is smaller than Earth's<br>D. Mars' centripetal acceleration is smaller than Earth's |
| Correct Answer: | D |
| **Question Type 3: Information-Based Questions** | |
| Question: | As shown in figure (a), the Chinese sci-fi movie "The Wandering Earth 2" depicts a space station built by humans in geostationary orbit, with a "space elevator" transporting people between Earth and the station. Figure (b) shows the relationship between the centripetal acceleration $a$ required due to Earth's rotation and the distance $r$ from the Earth's center. Given that $r_1$ is the Earth's radius and $r_2$ is the geosynchronous satellite orbital radius, which of the following statements is correct? |
| Figure: | 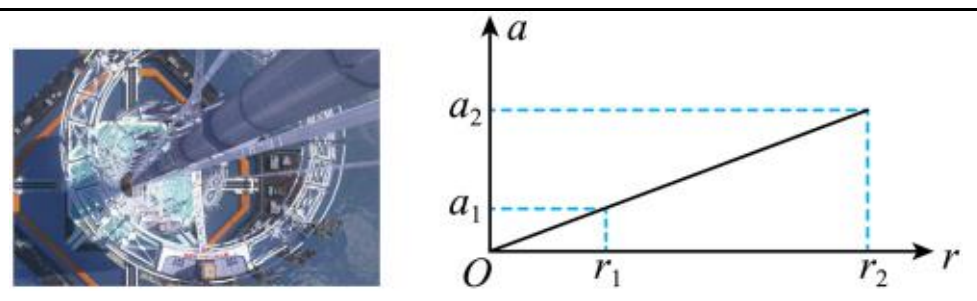（a） （b） |
| Options: | A . The angular velocity of Earth's rotation. $\omega = \frac{a_2 - a_1}{r_2 - r_1}$<br>B. The period of the geostationary satellite. $T = 2\pi\sqrt{\frac{r_2}{a_2}}$<br>C. The support force on a person by the elevator cabin remains constant during ascent.<br>D. Releasing an object freely from the space station will result in free-fall motion. |
| Correct Answer: | B |